\documentclass[floatfix,preprint,aps,showpacs]{revtex4}
\usepackage{epsfig}
\newcommand{\ee}{$e^{+}e^{-}$}
\newcommand{\NN}{$N$+$N$}
\newcommand{\nn}{$n$+$n$}
\newcommand{\CC}{$C$+$C$}
\newcommand{\np}{$n$+$p$}
\newcommand{\pp}{$p$+$p$}
\newcommand{\pd}{$p$+$d$}
\newcommand{\xdp}{$d$+$p$}
\newcommand{\gevu}{G$e$V/$u$}
\newcommand{\gevc}{G$e$V/$c^{2}$}

\begin{document}

\title{Origin of the low-mass electron pair excess in light nucleus-nucleus collisions}

\author{G.~Agakishiev$^{8}$, A.~Balanda$^{3}$,
D.~Belver$^{15}$, A.V.~Belyaev$^{6}$, A.~Blanco$^{2}$, M.~B\"{o}hmer$^{11}$, J.~L.~Boyard$^{13}$,
P.~Braun-Munzinger$^{4,b}$, P.~Cabanelas$^{15}$, E.~Castro$^{15}$, S.~Chernenko$^{6}$, T.~Christ$^{11}$, M.~Destefanis$^{8}$, J.~D\'{\i}az$^{16}$, F.~Dohrmann$^{5}$, A.~Dybczak$^{3}$, L.~Fabbietti$^{11}$, O.V.~Fateev$^{6}$, P.~Finocchiaro$^{1}$, P.~Fonte$^{2,a}$, J.~Friese$^{11}$, I.~Fr\"{o}hlich$^{7}$, T.~Galatyuk$^{7}$, J.~A.~Garz\'{o}n$^{15}$, R.~Gernh\"{a}user$^{11}$, A.~Gil$^{16}$, C.~Gilardi$^{8}$, M.~Golubeva$^{10}$, D.~Gonz\'{a}lez-D\'{\i}az$^{4}$, F.~Guber$^{10}$,
T.~Hennino$^{13}$, R.~Holzmann$^{4}$, I.~Iori$^{9,d}$, A.~Ivashkin$^{10}$, M.~Jurkovic$^{11}$,
B.~K\"{a}mpfer$^{5,c}$, T.~Karavicheva$^{10}$, D.~Kirschner$^{8}$, I.~Koenig$^{4}$, W.~Koenig$^{4}$,
B.~W.~Kolb$^{4}$, R.~Kotte$^{5}$, F.~Krizek$^{14}$, R.~Kr\"{u}cken$^{11}$, W.~K\"{u}hn$^{8}$,
A.~Kugler$^{14}$, A.~Kurepin$^{10}$, S.~Lang$^{4}$, J.~S.~Lange$^{8}$, K.~Lapidus$^{10}$,
T.~Liu$^{13}$, L.~Lopes$^{2}$, M.~Lorenz$^{7}$, L.~Maier$^{11}$, A.~Mangiarotti$^{2}$,
J.~Markert$^{7}$, V.~Metag$^{8}$, B.~Michalska$^{3}$, J.~Michel$^{7}$,
E.~Morini\`{e}re$^{13}$, J.~Mousa$^{12}$, C.~M\"{u}ntz$^{7}$, L.~Naumann$^{5}$, J.~Otwinowski$^{3}$,
Y.~C.~Pachmayer$^{7}$, M.~Palka$^{4}$, Y.~Parpottas$^{12}$, V.~Pechenov$^{4}$, O.~Pechenova$^{4}$,
J.~Pietraszko$^{4}$, W.~Przygoda$^{3}$, B.~Ramstein$^{13}$, A.~Reshetin$^{10}$, A.~Rustamov$^{4}$, A.~Sadovsky$^{10}$, P.~Salabura$^{3}$, A.~Schmah$^{11}$, E.~Schwab$^{4}$, Yu.G.~Sobolev$^{14}$, S.~Spataro$^{8}$, B.~Spruck$^{8}$, H.~Str\"{o}bele$^{7}$, J.~Stroth$^{7,4}$, C.~Sturm$^{7}$, M.~Sudol$^{13}$, A.~Tarantola$^{7}$, K.~Teilab$^{7}$, P.~Tlusty$^{14}$, M.~Traxler$^{4}$, R.~Trebacz$^{3}$, H.~Tsertos$^{12}$, V.~Wagner$^{14}$, M.~Weber$^{11}$, M.~Wisniowski$^{3}$, T.~Wojcik$^{3}$, J.~W\"{u}stenfeld$^{5}$, S.~Yurevich$^{4}$, Y.V.~Zanevsky$^{6}$, P.~Zhou$^{5}$}

\affiliation{
(HADES collaboration)
\\
\mbox{$^{1}$Istituto Nazionale di Fisica Nucleare - Laboratori Nazionali del Sud, 95125~Catania, Italy}\\
\mbox{$^{2}$LIP-Laborat\'{o}rio de Instrumenta\c{c}\~{a}o e F\'{\i}sica Experimental de Part\'{\i}culas,}\\
\mbox{3004-516~Coimbra, Portugal}\\
\mbox{$^{3}$Smoluchowski Institute of Physics, Jagiellonian University of Cracow, 30-059~Krak\'{o}w, Poland}\\
\mbox{$^{4}$GSI Helmholtzzentrum f{\"u}r Schwerionenforschung GmbH, 64291~Darmstadt, Germany}\\
\mbox{$^{5}$Institut f\"{u}r Strahlenphysik, Forschungszentrum Dresden-Rossendorf, 01314~Dresden, Germany}\\
\mbox{$^{6}$Joint Institute of Nuclear Research, 141980~Dubna, Russia}\\
\mbox{$^{7}$Institut f\"{u}r Kernphysik, Goethe-Universit\"{a}t, 60438 ~Frankfurt, Germany}\\
\mbox{$^{8}$II.Physikalisches Institut, Justus Liebig Universit\"{a}t Giessen, 35392~Giessen, Germany}\\
\mbox{$^{9}$Istituto Nazionale di Fisica Nucleare, Sezione di Milano, 20133~Milano, Italy}\\
\mbox{$^{10}$Institute for Nuclear Research, Russian Academy of Science, 117312~Moscow, Russia}\\
\mbox{$^{11}$Physik Department E12, Technische Universit\"{a}t M\"{u}nchen, 85748~M\"{u}nchen, Germany}\\
\mbox{$^{12}$Department of Physics, University of Cyprus, 1678~Nicosia, Cyprus}\\
\mbox{$^{13}$Institut de Physique Nucl\'{e}aire (UMR 8608),\\ CNRS/IN2P3 - Universit\'{e} Paris Sud,}\\
\mbox{F-91406~Orsay Cedex, France}\\
\mbox{$^{14}$Nuclear Physics Institute, Academy of Sciences of Czech Republic, 25068~Rez, Czech Republic}\\
\mbox{$^{15}$Departamento de F\'{\i}sica de Part\'{\i}culas, Univ. de Santiago de Compostela,}\\
\mbox{15706~Santiago de Compostela, Spain}\\
\mbox{$^{16}$Instituto de F\'{\i}sica Corpuscular, Universidad de Valencia-CSIC, 46971~Valencia, Spain}\\
\\
\mbox{$^{a}$ also at ISEC Coimbra,~Coimbra, Portugal}\\
\mbox{$^{b}$ also at ExtreMe Matter Institute EMMI, 64291 Darmstadt, Germany}\\
\mbox{$^{c}$ also at Technische Universit\"{a}t Dresden, 01062~Dresden, Germany}\\
\mbox{$^{d}$ also at Dipartimento di Fisica, Universit\`{a} di Milano, 20133~Milano, Italy}\\
}

\begin{abstract}
We report measurements of electron pair production in elementary \pp\ and \xdp\ reactions at $1.25$~\gevu\ with the HADES spectrometer. For the first time, the electron pairs were reconstructed for \np\ reactions by detecting the proton spectator from the deuteron breakup. We find that the yield of electron pairs with invariant mass $\textrm{M}_{e^+e^-} > 0.15$~\gevc\ is about an order of magnitude larger in \np\ reactions as compared to \pp. A comparison to model calculations demonstrates that the production mechanism is not sufficiently described yet. The electron pair spectra measured in \CC\ reactions are compatible with a superposition of elementary \np\ and \pp\ collisions, leaving little room for additional electron pair sources in such light collision systems.
\end{abstract}
\maketitle
The formation and investigation of strongly interacting matter at high temperature and density is in the focus of experiments with relativistic and ultra-relativistic heavy-ion beams. In recent years, dilepton spectroscopy has been established as a valuable tool to probe such extreme matter states. Experiments performed at SPS ($40-158$~\gevu) and RHIC ($\sqrt{s_{NN}}=200$~G$e$V) energies found a significant excess of lepton pairs originating from the hot and dense phase of the formed matter, over-shining contributions from electromagnetic decays of long-lived mesons during the later stages of the collisions~\cite{ceres_na60_phenix}.
At lower beam energies ($1-2$~\gevu) electron pair (\ee) production has been studied by DLS~\cite{dls} and, more recently, by HADES~\cite{hades_tech}. Even for the light collision system \CC\ a significant electron pair excess above long-lived sources was identified in the invariant mass range of $0.15 < \textrm{M}_{e^+e^-} / (\textrm{G}e\textrm{V}/c^2) < 0.6$~\cite{hades_prl,hades_plb}.
The new HADES results confirm remarkably well the DLS data~\cite{hades_plb}, which could not satisfactorily be explained by various transport models for more than a decade (for a review see~\cite{cassing99}).
However, in contrast to the situation at high beam energies, the question could not be answered whether the observed excess is related to the onset of in-medium effects and not to some insufficiently described elementary dilepton sources. This dilemma can be traced back to the quite different composition of the strongly interacting matter formed at these low energies where baryons, mainly nucleons ($N$) and $\Delta(1232)$ resonances, dominate over mesonic degrees of freedom~\cite{metag}. A fully microscopic description, however, suffers from poorly known elementary processes like Dalitz decays of baryonic resonances (i.e.\ $\Delta,N^*\rightarrow Ne^+e^-$) and non-resonant $NN$ bremsstrahlung.


By comparing such model calculations \cite{bratkovskaya,shekhter,ernst} with data obtained by DLS for \pp\ and \pd\ collisions at energies near the $\eta$ production threshold ($\textrm{E}_{\mathrm{beam}}=1.27$~\gevu) in \NN\ collisions~\cite{wilson} it has been concluded that the electron pair yield can qualitatively be understood assuming three sources: (i) $\pi^0$ Dalitz decay, (ii) $\Delta$ Dalitz decay ($\Delta \rightarrow N e^+ e^-$), and (iii) "quasi-elastic" \NN\ scattering $NN\rightarrow NNe^+e^-$ (bremsstrahlung). OBE model calculations~\cite{kaptari,shyam,jong} show that the bremsstrahlung and the $\Delta$ Dalitz contributions appear to be almost equally important for \np\ collisions, while for \pp\ collisions the $\Delta^+$ decay plays the dominant role and bremsstrahlung is strongly suppressed. However, the calculations differ in the absolute cross sections; at $1.04$~G$e$V for example, results of a recent work~\cite{kaptari} reveal a $\sim 2-4$ times larger yield (depending on mass) compared to other calculations. This new outcome has triggered a series of microscopic transport calculations~\cite{bratkovskaya} which are successful in explaining the pair spectra measured in \CC\ collisions by DLS as well as by HADES.

The aim of the experiments reported in the following was to further constrain this still not conclusive interpretation. Two dedicated experimental runs were performed with the High Acceptance Di-Electron Spectrometer~\cite{hades_tech} installed at the GSI Helmholtzzentrum f{\"u}r Schwerionenforschung, Germany. Proton and deuteron beams of $10^7$ particles/s with kinetic energies of $1.25$~\gevu\ were incident on a liquid hydrogen cell with a length of $5$~cm, corresponding to a total thickness of $\rho d=0.35$~g/cm$^2$. Quasi-free \np\ reactions were selected on trigger level by detection of fast spectator protons from the deuterium break-up in a dedicated Forward hodoscope Wall (FW)~\cite{lapidus}. It was placed $7$~m downstream of the target and covered polar angles between $0.3^\circ$ and $7^\circ$. Charged particles ($p,~\pi^\pm,~e^\pm$) were detected in the spectrometer as described in~\cite{hades_tech}.

In \pp\ reactions, the data readout was started upon a first-level trigger (LVL1) decision with two different settings requiring: (LVL1A) a charged-particle multiplicity $\textrm{MUL} \geq 3$ in HADES or (LVL1B) $\textrm{MUL} \geq 2$ with hits in opposite sectors of the time-of-flight detectors. These two trigger conditions were chosen to enrich inclusive electron pair production ($pp\rightarrow e^+e^-X$) and elastic \pp\ scattering for normalization purposes, respectively. The trigger efficiency of LVL1A was studied in Monte Carlo simulations. It is nearly independent of the pair mass and amounts to 0.84. LVL1 was followed by a second-level trigger (LVL2)~\cite{hades_tech} requesting at least one electron track candidate. All events with a positive LVL2 decision and every fifth LVL1 event, disregarding the LVL2 decision, were written to tape (in total $7.9 \cdot 10^{8}$ events). Electron identification, track reconstruction, and electron pair (unlike- and like-sign) reconstruction were performed as described in detail in~\cite{hades_prl,hades_plb,hades_tech}. The combinatorial background (CB) was obtained from same-event like-sign pairs using the arithmetic mean
$d\textrm{N}^{CB}/d\textrm{M}_{e^+e^-} \equiv (d\textrm{N}/d\textrm{M}_{e^-e^-} + d\textrm{N}/d\textrm{M}_{e^+e^+})$
to account for correlated background from double conversion of $\pi^0$ decay photons or conversion of the photon accompanying Dalitz decays, as well as for uncorrelated $e^+e^-$ stemming from multi-pion decays. The final invariant mass distribution of signal pairs is obtained by subtracting the CB from the corresponding unlike-sign pair distribution corrected, pair by pair, for the detector and reconstruction inefficiencies~\cite{hades_tech}. In total $39 \cdot 10^3$ signal pairs, $\sim 350$ hereof in the region above $0.15$~\gevc\ with a signal-to-CB ratio $\geq 1$~\cite{przygoda}, were reconstructed.

\begin{figure}[tb]
   \mbox{\epsfig{figure={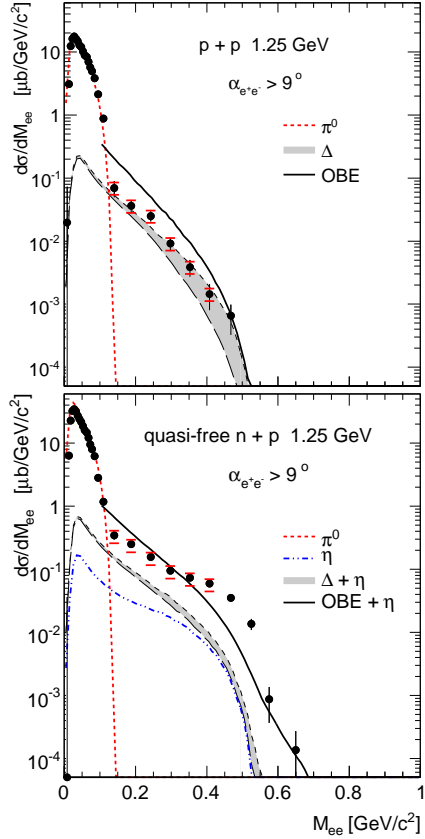}, width=0.5\linewidth}}
   \vspace*{-0.2cm}
   \caption[]{(Color online) Electron pair differential cross sections as function of invariant mass (full circles) measured in \pp\ reactions (upper) and in quasi-free \np\ reactions (lower panel) at $1.25$~G$e$V. Systematic errors (constant in the whole mass range) are indicated by (red) horizontal bars, statistical errors by vertical bars. In the analysis, \ee\ pairs with an opening angle of $\alpha \leq 9^{\circ}$ are removed from the sample. The lines show results of model calculations with the Pluto event generator (see text for explanations).
   }
\label{fig1_new}
\vspace*{-0.3cm}
\end{figure}
The inclusive cross section for electron pair production in \pp\ collisions as a function of the pair invariant mass is shown in Fig.~\ref{fig1_new} (upper panel). The measured pair yield was normalized to the \pp\ elastic scattering yield, corrected for reconstruction and trigger inefficiencies, and multiplied by the known differential elastic cross sections in the acceptance~\cite{edda}.
The overall normalization error of this procedure is estimated to be $9\%$ and does not show any pair-mass dependence. It results from the error on the published elastic cross section ($5\%$) and from systematic errors related to the reconstruction of elastic-scattering events ($7\%$).
An additional uncorrelated systematic uncertainty of $20\%$ comes from the pair reconstruction efficiency. It can include a smooth invariant mass dependence and is added in quadrature.

The data are compared to simulated pair distributions calculated with the Pluto event generator~\cite{pluto2} assuming essentially $\pi^0$ and $\Delta^+$ Dalitz decays (see Fig.~\ref{fig1_new}). The measured yield in the $\pi^0$ Dalitz decay region is reproduced taking into account the inclusive $\pi^0$ production cross section ($\sigma_{\pi^0}^{pp}=4.5\pm0.9$ mb) from the resonance model~\cite{teis}, which describes the existing data~\cite{bystricky}, and the measured $\pi^0\rightarrow e^+e^-\gamma$ branching ratio ($1.2\pm0.032 \%$~\cite{pdg}).
To model the emission rate in the mass region above the $\pi^0$-Dalitz region we follow the procedure used in microscopic transport calculations. Since at an energy of $1.25$~\gevu\ pions are produced mostly through intermediate $\Delta$ resonances~\cite{teis}, the respective cross section for $\Delta^+$ production has been fixed to  $\sigma_{\Delta^+}=3/2\; \sigma_{\pi^0}$. As discussed in~\cite{krivoruchenko}, there are different prescriptions for the differential partial decay width
$d\Gamma_{\Delta \rightarrow Ne^+ e^-}(\textrm{M}_{e^+e^-})/d{\textrm{M}_{e^+e^-}}$.
In a quark-model picture the $\Delta$ radiative decay can be associated with a spin flip and pure S-wave states for the quarks. Such a magnetic dipole transition is fully described by a magnetic transition form factor ($\textrm{G}_\mathrm{M}$) and its magnitude at the photon point $\textrm{G}_\mathrm{M}(0) = 3.02\pm0.03$, extracted from pion photo-production experiments~\cite{pascalutsa}, is reproduced by~\cite{krivoruchenko}. In our simulation~\cite{pluto2} we hence set electric and Coulomb transition form factors to zero, and use the expression for the $\Delta$ Dalitz decay differential width given in~\cite{krivoruchenko}; the result is shown in the upper panel of Fig.~\ref{fig1_new} (long dashed line). In this approach, a possible modification of the magnetic transition form factor due to intermediate vector mesons is not treated and it can therefore be regarded as a lower bound for $\Delta$ Dalitz contributions to the pair spectrum. To illustrate the variation in pair yield due to different prescriptions of the form factor we also include in Fig.~\ref{fig1_new} (short dashed line) the result of a calculation using the two-component quark model~\cite{iachello}, which is mostly driven in our kinematical range by Vector Meson Dominance (VMD). As expected, an enhanced yield is observed, in particular for high pair masses. Note that this model seems to provide a better description of the \pp\ data.

Next we also include the predictions of the OBE model calculations discussed above. We have parameterized the calculated differential cross sections obtained in~\cite{kaptari}; we have further assumed isotropic virtual photon emission and have included corrections due to \NN\ final state interactions. Details of the implementation can be found in~\cite{pluto2}. The result of the simulation is shown in Fig.~\ref{fig1_new} as solid black line. The yield calculated in this approach overestimates the measured spectrum.

%
\begin{figure}[tb]
   \mbox{\epsfig{figure={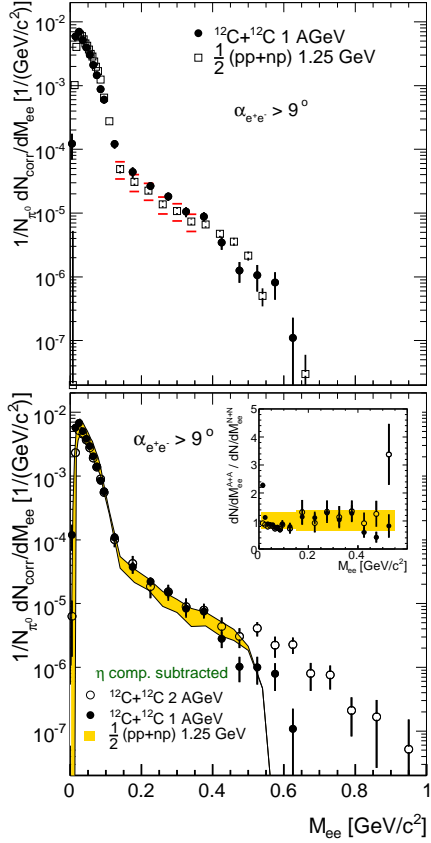}, width=0.5\linewidth}}
   \vspace*{-0.2cm}
   \caption[]{(Color online) Upper panel: Electron pair yield as a function of the invariant mass measured in \CC\ collisions (full circles) at $1$~\gevu\ compared to the reference yield obtained from \pp\ and \np\ collisions (open squares, errors as defined in Fig.~\ref{fig1_new}). Lower panel: Comparison of the reference spectrum from elementary collisions with the HADES results for \CC\ collisions at $1$ (full circles) and $2$~\gevu\ (open circles). Contributions from  $\eta$ Dalitz decay have been subtracted here. The yellow band exactly corresponds to the systematical errors.
   The inset displays the ratio to the reference spectrum in the interesting mass range. All distributions are normalized to the corresponding number of produced $\pi^0$ mesons.}
\vspace*{-0.3cm}
\label{fig2_new}
\end{figure}

We now discuss the deuteron induced quasi-free \np\ reactions. The running conditions were the same as the ones used for the \pp\ run, except that LVL1A also required a coincidence with at least one charged particle hit in FW. In total, $1.3 \cdot 10^{9}$ events were recorded for \xdp\ reactions. The lower panel of Fig.~\ref{fig1_new} displays the inclusive cross section for electron pair production measured in coincidence with the spectator proton in FW. To enhance the spectator character of the forward detected proton and suppress other reaction types we imposed a condition on its momentum ($1.6 < p_{sp} / (\textrm{G}e\textrm{V}/c) <2.6$). The moderate experimental momentum resolution obtained from a time-of-flight measurement in FW enforces the given range of $p_{sp}$. As for the \pp\ reactions, the dielectron invariant mass spectrum has been corrected for all inefficiencies and the CB has been subtracted. The overall normalization is obtained in an analogous way as for the \pp\ reactions using the simultaneously measured $\textrm{(quasi-)elastic}$ \pp\ scattering yield. The total statistics of signal pairs amounts to $36 \cdot 10^3$ and to $1454$ for pairs with invariant mass $\textrm{M}_{e^+e^-}>0.15$~\gevc.

The pair cross section in the $\pi^0$ mass region
is a factor of $\sim 2$ larger as compared to the \pp\ reaction, in accordance with the prediction of the resonance model~\cite{teis}. The good agreement between the measured and the simulated yield in the $\pi^0$ mass region confirms our analysis and normalization procedure. The shape of the mass spectra changes dramatically when going from \pp\ to \np\ interactions. In the intermediate mass region ($0.15$ to $0.35$~\gevc) the \np\ yield is enhanced by a factor of about ten over the \pp\ yield while one would expect only a factor two if the $\Delta$ were the only relevant source. Furthermore, in \np\ reactions, the tail at high invariant mass extends much further and the ratio of the two spectra reaches almost a value of $100$ at $0.5$~\gevc.
A similar observation was also made by DLS in \pd\ experiments for which the quasi-free \np\ reactions could however not be isolated~\cite{wilson}. To further test the validity of the spectator assumption we studied the shape of the pair spectrum restricting the spectator emission angle to a very forward cone ($0.3^{\circ} \le \theta_{sp} \le 2^{\circ}$). No significant change of the shape of the resulting pair spectrum was observed~\cite{lapidus}.

To model the \np\ data we proceeded as in the \pp\ case, but added the following features to the simulation:
(i) the available energy in the center of mass was smeared to include the neutron momentum distribution in the deuteron using the Paris potential~\cite{cosy_tof} and (ii) contributions from $\eta$ Dalitz decays were accounted for (dashed-double dotted line). The cross sections for $np\rightarrow np\eta$ and $np\rightarrow d\eta$ reactions are known down to the production threshold (for a review see~\cite{cosy}). As it can be seen from Fig.~\ref{fig1_new}, the cocktail obtained in this way does not account for the measured yield. Moreover, also the result of the OBE calculation~\cite{kaptari} (black solid line) does not describe the data (black solid circles) and, in contrast to the \pp\ reactions, underestimates the observed yield in the high mass region.
The data show an enhanced emission in the high mass region, well beyond contributions from $\eta$ Dalitz decay, while the calculation follows roughly an exponential slope. Note that the $\eta$ contribution has been added to the model calculations for both cases.

Leaving this interesting question to further investigations, we continue by comparing the dielectron invariant mass distributions measured in \CC\ reactions to a superposition of the yields measured in the elementary \NN\ collisions. In order to obtain the latter we assumed that electron pair production in \nn\ collisions is the same as in \pp, i.e.\ we considered the elastic bremsstrahlung process being small and use as \NN\ reference spectrum the dielectron yield defined by
$\sigma_{NN}(\textrm{M}_{e^+e^-})=0.5\cdot(\sigma_{pp}(\textrm{M}_{e^+e^-})+\sigma_{pn}(\textrm{M}_{e^+e^-}))$.
In~Fig.~\ref{fig2_new} (upper part) we compare the reference spectrum to the dielectron yield obtained previously in \CC\ at 1~\gevu~\cite{hades_plb}. To be consistent, we converted the dielectron cross section to relative multiplicity according to
$1/\textrm{N}_{\pi^0} \cdot d\textrm{N}/d\textrm{M}_{e^+e^-} = 1/\sigma_{\pi^0} \cdot d\sigma/d\textrm{M}_{e^+e^-},$
where $\sigma_{\pi^0}$ was taken from our measurements~\cite{hades_pion}. Both distributions agree well over the full mass range, though measured at slightly different beam energies. We like to emphasize that the energy dependence is taken out to some extent due to the normalization to neutral pion production; it is known that the excess electron pair yields observed in \CC\ collisions above contributions from long-lived sources exhibit a scaling with beam energy like pion production~\cite{hades_plb}. Furthermore, the contribution from $\eta$ Dalitz decay in the relevant mass range is small, i.e.\ $<15\%$. The absence of a strong beam energy dependence is demonstrated in the lower panel of Fig.~\ref{fig2_new}, where the \NN\ reference spectrum is compared with both our \CC\ results at $1$ and $2$~\gevu. To better visualize the scaling behavior of the excess yield over long-lived sources, the $\eta$ contribution has been subtracted using the data from~\cite{taps,cosy}. A very good agreement between all collision systems can be observed in the excess region ($0.15 < \textrm{M}_{e^+e^-} / (\textrm{G}e\textrm{V}/c^2) < 0.5$), suggesting a common source for the excess pairs, scaling with beam energy like pion production (see insert in Fig.~\ref{fig2_new}). The reduced phase space at the lower beam energies affects evidently the high-mass region ($\textrm{M}_{e^+e^-}>0.5$~\gevc) only. To correct for a slight difference in acceptance between the runs with carbon target and LH$_{2}$ target the reference spectrum was scaled up by a factor of $1.28$. Hence we conclude that the so-called ''DLS puzzle'', i.e.\ the so far unexplained excess of electron pairs above contributions from long-lived sources, has its origin in a hitherto insufficient treatment of radiation from elementary \NN\ collisions.

In summary, we have measured dielectron production in \pp\ and quasi-free \np\ collisions at $1.25$~G$e$V. A very strong isospin dependence of the dielectron production has been found. We have shown that the puzzling dielectron excess in the intermediate mass range of $0.15<\textrm{M}_{e^+e^-}/(\textrm{G}e\textrm{V}/c^2) <0.5$ observed in \CC\ collisions at $1$ and $2$~\gevu\ can be described by a superposition of elementary \pp\ and \np\ collisions. Although a sound theoretical description of the relevant sources is still lacking, the excess can be traced back essentially to effects present already in \np\ collisions. Further investigations to search for significant medium effects, based on the reference established in this work, are planned by HADES and will concentrate on heavier collision systems.

The collaboration gratefully acknowledges the support by BMBF grants 06TM970I, 06GI146I, 06F-140, and 06DR135 (Germany), by GSI (TM-FR1,GI/ME3,OF/STR), by Helmholtz Alliance HA216/EMMI, by grants GA CR 202/00/1668 and GA AS CR IAA1048304 (Czech Republic), by grant KBN 1P03B 056 29 (Poland), by INFN (Italy), by CNRS/IN2P3 (France), by grants MCYT FPA2000-2041-C02-02 and XUGA PGID T02PXIC20605PN (Spain), by grant UCY-10.3.11.12 (Cyprus), by INTAS grant 03-51-3208 and by EU contract RII3-CT-2004-506078.
%

%
\end{document}